\documentclass[onecolumn,showpacs,11pt]{revtex4}

\usepackage{graphicx}% Include figure files
\usepackage{dcolumn}% Align table columns on decimal point
\usepackage{bm}% bold math

\input epsf

\begin{document}

\title{\Large Constraining the Parameters of Modified Chaplygin Gas in Einstein-Aether Gravity}

\author{\bf Ujjal Debnath\footnote{ujjaldebnath@yahoo.com ,
ujjal@iucaa.ernet.in}}

\affiliation{Department of Mathematics, Bengal Engineering and
Science University, Shibpur, Howrah-711 103, India.\\}

\date{\today}

\begin{abstract}
We have assumed FRW model of the universe in Einstein-Aether
gravity filled with dark matter and Modified Chaplygin gas (MCG)
type dark energy. We present the Hubble parameter in terms of some
unknown parameters and observational parameters with the redshift
$z$. From observed Hubble data (OHD) set (12 points), we have
obtained the bounds of the arbitrary parameters ($A,B$) of MCG by
minimizing the $\chi^{2}$ test. Next due to joint analysis of BAO
and CMB observations, we have also obtained the best fit values
and the bounds of the parameters ($A,B$) by fixing some other
parameters. We have also taken type Ia supernovae data set (union
2 data set with 557 data points). Next due to joint analysis with
SNe, we have obtained the best fit values of parameters. The
best-fit values and bounds of the parameters are obtained by 66\%,
90\% and 99\% confidence levels for OHD, OHD+BAO, OHD+BAO+CMB and
OHD+BAO+CMB+SNe joint analysis. The distance modulus $\mu(z)$
against redshift $z$ for our theoretical MCG model in
Einstein-Aether gravity have been tested for the best fit values
of the parameters and the observed SNe Ia union2 data sample.
\end{abstract}


\maketitle

\section{\normalsize\bf{Introduction}}

Observational evidence strongly points to an accelerated expansion
of the Universe, but the physical origin of this acceleration is
unknown. The observations include type Ia Supernovae and Cosmic
Microwave Background (CMB)
\cite{Perlmutter,Riess,Riess1,Bennet,Sperge} radiation. The
standard explanation invokes an unknown ``dark energy'' component
which has the property that positive energy density and negative
pressure. Observations indicate that dark energy occupies about
70\% of the total energy of the universe, and the contribution of
dark matter is $\sim$ 26\%. This accelerated expansion of the
universe has also been strongly confirmed by some other
independent experiments like Sloan Digital Sky Survey (SDSS)
\cite{Adel}, Baryonic Acoustic Oscillation (BAO)
\cite{Eisenstein}, WMAP data analysis \cite{Briddle,Spergel} etc.
Over the past decade there have been many theoretical models for
mimicking the dark energy behaviors, such as the simplest (just)
cosmological constant in which the equation of state is
independent of the cosmic time and which can fit the observations
well. This model is the so-called $\Lambda$CDM, containing a
mixture of cosmological constant $\Lambda$ and cold dark matter
(CDM). However, two problems arise from this scenario, namely
``fine-tuning'' and the ``cosmic coincidence'' problems. In order
to solve these two problems, many dynamical dark energy models
were suggested, whose equation of state evolves with cosmic time.
The scalar field or quintessence \cite{Peebles,Cald} is one of the
most favored candidate of dark energy which produce sufficient
negative pressure to drive acceleration. In order to alleviate the
cosmological-constant problems and explain the acceleration
expansion, many dynamical dark energy models have been proposed,
such as K-essence, Tachyon, Phantom, quintom, Chaplygin gas model,
etc \cite{Arme,Sen,Cald1,Feng,Kamen}. Also the interacting dark
energy models including Modified Chaplygin gas \cite{Debnath},
holographic dark energy model \cite{Cohen}, and braneworld model
\cite{Sahni} have been proposed. Recently, based on principle of
quantum gravity, the agegraphic dark energy (ADE) and the new
agegraphic dark energy (NADE) models were proposed by Cai
\cite{Cai} and Wei et al \cite{Wei} respectively. The theoretical
models have been tally with the observations with different data
sets say TORNY, Gold sample data sets
\cite{Paddy1,Riess1,Tonry,Barris}. In Einstein's gravity, the
modified Chaplygin gas \cite{Debnath} best fits with the 3 year
WMAP and the SDSS data with the choice of parameters $A =0.085$
and $\alpha = 1.724$ \cite{Lu} which are improved constraints than
the previous
ones $-0.35 < A < 0.025$ \cite{Jun}.\\

Another possibility is that general relativity is only accurate on
small scales and has to be modified on cosmological distances. One
of these is a modified gravity theories. In this case cosmic
acceleration would arise not from dark energy as a substance but
rather from the dynamics of modified gravity. Modified gravity
constitutes an interesting dynamical alternative to $\Lambda$CDM
cosmology in that it is also able to describe the current
acceleration in the expansion of our universe. The simplest
modified gravity is DGP brane-world model \cite{Dvali}. The other
alternative approach dealing with the acceleration problem of the
Universe is changing the gravity law through the modification of
action of gravity by means of using $f(R)$ gravity \cite{An}
instead of the Einstein-Hilbert action. Some of these models, such
as $1/R$ and logarithmic models, provide an acceleration for the
Universe at the present time \cite{clif}. Other modified gravity
includes $f(T)$ gravity, $f(G)$ gravity, Gauss-Bonnet gravity,
Horava-Lifshitz gravity, Brans-Dicke gravity, etc
\cite{Yer,Noj,An1,Hora,Brans}.\\

In the present work, we concentrate on the generalized
Einstein-Aether theories as proposed by Zlosnik et al
\cite{Zlos,Zlos1}, which is a generalization of the
Einstein-Aether theory developed by Jacobson et al
\cite{Jacob,Jacob1}. These years a lot of work has been done in
generalized Einstein-aether theories
\cite{Gar,Linder,Barrow,Junt,Li,Gasp1,Gasp2}. In the generalized
Einstein-Aether theories by taking a special form of the
Lagrangian density of Aether field, the possibility of
Einstein-Aether theory as an alternative to dark energy model is
discussed in detail, that is, taking a special Aether field as a
dark energy candidate and it has been found the constraints from
observational data \cite{Meng1,Meng2}. Since modified gravity
theory may be treated as alternative to dark energy, so Meng et al
\cite{Meng1,Meng2} have not taken by hand any types of dark energy
in Einstein-Aether gravity and shown that the gravity may be
generates dark energy. Here if we exempt this assumption, so we
need to consider the dark energy from outside. So we assume the
FRW universe in Einstein-Aether gravity model filled with the dark
matter and the modified Chaplygin gas (MCG) type dark energy. The
basic concepts of Einstein-Aether gravity theory are presented in
section II. The modified Friedmann equations and their solutions
are given in section III. The observational data analysis tools in
observed Hubble data (OHD), OHD+BAO, OHD+BAO+CMB and
OHD+BAO+CMB+SNe for $\chi^{2}$ minimum test will be studied in
section IV and investigate the bounds of unknown parameters
$(A,B)$ of MCG dark energy by fixing other parameters. The
best-fit values of the parameters are obtained by 66\%, 90\% and
99\% confidence levels. The distance modulus $\mu(z)$ against
redshift $z$ for our theoretical model of the MCG in
Einstein-Aether gravity model for the best fit values of the
parameters and the observed SNe Ia union2 data sample. Finally we
present the conclusions of the work in section V.

\section{\normalsize\bf{Einstein-Aether Gravity Theory}}

In order to include Lorentz symmetry violating terms in
gravitation theories, apart from some noncommutative gravity
models, one may consider existence of preferred frames. This can
be achieved admitting a unit timelike vector field in addition to
the metric tensor of spacetime. Such a timelike vector implies a
preferred direction at each point of spacetime. Here the unit
timelike vector field is called the {\it Aether} and the theory
coupling the metric and unit timelike vector is called the {\it
Einstein-Aether} theory \cite{Jacob}. So Einstein-Aether theory is
the extension of general relativity (GR) that incorporates a
dynamical unit timelike vector field (i.e., Aether). In the last
decade there is an increasing interest in the Aether theory.\\

The action of the Einstein-Aether gravity theory with the normal
Einstein-Hilbert part action can be written in the form
\cite{Zlos,Meng1}
\begin{equation}
S=\int d^{4}x\sqrt{-g}\left[\frac{R}{16\pi G}+{\cal L}_{EA}+{\cal
L}_{m} \right]
\end{equation}

where ${\cal L}_{EA}$ is the vector field Lagrangian density while
${\cal L}_{m}$ denotes the Lagrangian density for all other matter
fields. The Lagrangian density for the vector part consists of
terms quadratic in the field \cite{Zlos,Meng1}:

\begin{equation}
{\cal L}_{EA}=\frac{M^{2}}{16\pi G}~F(K)+\frac{1}{16\pi
G}~\lambda(A^{a}A_{a}+1)~,
\end{equation}
\begin{equation}
K=M^{-2}{K^{ab}}_{cd}\nabla_{a}A^{c}\nabla_{b}A^{d}~,
\end{equation}
\begin{equation}
{K^{ab}}_{cd}=c_{1}g^{ab}g_{cd}+c_{2}\delta^{a}_{c}\delta^{b}_{d}+c_{3}\delta^{a}_{d}\delta^{b}_{c}
\end{equation}

where $c_{i}$ are dimensionless constants, $M$ is the coupling
constant which has the dimension of mass, $\lambda$ is a Lagrange
multiplier that enforces the unit constraint for the time-like
vector field, $A^{a}$ is a contravariant vector, $g_{ab}$ is
metric tensor and $F(K)$ ia an arbitrary function of $K$. From
(1), we get the field equations
\begin{equation}
G_{ab}=T_{ab}^{EA}+8\pi G T_{ab}^{m}~,
\end{equation}
\begin{equation}
\nabla_{a}\left(F'{J^{a}}_{b}\right)=2\lambda A_{b}
\end{equation}
where
\begin{equation}
F'=\frac{dF}{dK}~~and~~{J^{a}}_{b}=2{K^{ad}}_{bc}\nabla_{d}A^{c}
\end{equation}
Here $T_{ab}^{m}$ is the energy momentum tensor for matter field
and $T_{ab}^{EA}$ is the energy momentum tensor for the vector
field and they are respectively given as follows: \cite{Meng1}
\begin{equation}
T_{ab}^{m}=(\rho+p)u_{a}u_{b}+pg_{ab}
\end{equation}
where $\rho$ and $p$ are respectively the energy density and
pressure of matter and $u_{a}=(1,0,0,0)$ is the fluid 4-velocity
vector and
\begin{equation}
T_{ab}^{EA}=\frac{1}{2}~\nabla_{d}\left[
\left({J_{(a}}^{d}A_{b)}-{J^{d}}_{(a}A_{b)}-J_{(ab)}A^{d}
\right)F' \right]-Y_{(ab)}F'+\frac{1}{2}~g_{ab}M^{2}F+\lambda
A_{a}A_{b}
\end{equation}
with
\begin{equation}
Y_{ab}=-c_{1}\left[
(\nabla_{d}A_{a})(\nabla^{d}A_{b})-(\nabla_{a}A_{d})(\nabla_{b}A^{d})
\right]
\end{equation}
where the subscript $(ab)$ means symmetric with respect to the
indices involved and $A^{a}=(1,0,0,0)$ is non-vanishing time-like
unit vector satisfying $A^{a}A_{a}=-1$.\\

\section{\normalsize\bf{Modified Friedmann Equations and Solutions}}

We consider the Friedmann-Robertson-Walker (FRW) metric of the
universe as
\begin{equation}
ds^{2}=-dt^{2}+a^{2}(t)\left[\frac{dr^{2}}{1-kr^{2}}+r^{2}\left(d\theta^{2}+sin^{2}\theta
d\phi^{2}\right) \right]
\end{equation}
where $k~(=0,\pm 1)$ is the curvature scalar and $a(t)$ is the
scale factor. From equations (3) and (4), we get
\begin{equation}
K=M^{-2}\left(c_{1}g^{ab}g_{cd}+c_{2}\delta^{a}_{c}\delta^{b}_{d}+c_{3}
\delta^{a}_{d}\delta^{b}_{c}\right)=\frac{3\beta H^{2}}{M^{2}}
\end{equation}
where $\beta=c_{1}+3c_{2}+c_{3}$ is constant. From eq. (5), we get
the modified Friedmann equation for Einstein-Aether gravity as in
the following \cite{Zlos,Meng1}:
\begin{equation}
\beta\left(-F'+\frac{F}{2K}\right)H^{2}+\left(H^{2}+\frac{k}{a^{2}}\right)=\frac{8\pi
G}{3}~\rho
\end{equation}
and
\begin{equation}
\beta\frac{d}{dt}\left(HF'\right)+\left(-2\dot{H}+\frac{2k}{a^{2}}\right)=8\pi
G(\rho+p)
\end{equation}
where $H~(=\frac{\dot{a}}{a})$ is Hubble parameter. Now we see
that if the first expressions of L.H.S. of equations (12) and (13)
are zero, we get the usual field equations for Einstein's gravity.
So first expressions arise for Einstein-Aether gravity. Also the
conservation equation is given by
\begin{equation}
\dot{\rho}+3\frac{\dot{a}}{a}(\rho+p)=0
\end{equation}

Now, assume that the matter fluid is combination of dark matter
and modified Chaplygin gas type dark energy. So
$\rho=\rho_{m}+\rho_{ch}$ and $p=p_{m}+p_{ch}$, where $\rho_{m}$
and $p_{m}$ are respectively the energy density and pressure of
dark matter and $\rho_{ch}$ and $p_{ch}$ are respectively the
energy density and pressure of modified Chaplygin gas. Assume that
the dark matter follows the barotropic equation of state
$p_{m}=w_{m}\rho_{m}$, where $w_{m}$ is a constant. The equation
of state of modified Chaplygin gas (MCG) is given by
\cite{Debnath}
\begin{equation}
p_{ch}=A\rho_{ch}-\frac{B}{\rho_{ch}^{\alpha}}
\end{equation}

where $A>0$, $B>0$ and $0\le\alpha\le 1$. Now we assume that there
is no interaction between dark matter and dark energy. So they are
separately conserved. From equation (14), we obtain the
conservation equations for dark matter and dark energy in the
form:
\begin{equation}
\dot{\rho}_{m}+3\frac{\dot{a}}{a}(\rho_{m}+p_{m})=0~~~~~and~~~~~
\dot{\rho}_{ch}+3\frac{\dot{a}}{a}(\rho_{ch}+p_{ch})=0
\end{equation}

Using equation of states and the conservation equations (17), we
obtain $\rho_{m}=\rho_{m0}(1+z)^{3(1+w_{m})}$ and
\begin{equation}
\rho_{ch}=\left[\frac{B}{1+A}+C(1+z)^{3(1+A)(1+\alpha)}
\right]^{\frac{1}{1+\alpha}}
\end{equation}

where $\rho_{m0}$ and $C$ are positive constants in which
$\rho_{m0}$ represents the present value of the density of dark
matter and $z=\frac{1}{a}-1$ is the cosmological redshift
(choosing $a_{0}=1$). The above expression can be written in the
form:
\begin{equation}
\rho_{ch}=\rho_{ch0}\left[A_{s}+(1-A_{s})(1+z)^{3(1+A)(1+\alpha)}
\right]^{\frac{1}{1+\alpha}}
\end{equation}

where $\rho_{ch0}$ is the present value of the MCG density and
$A_{s}=\frac{B}{(1+A)C+B}$. So $0\le A_{s}\le 1$.\\

Now since $F(K)$ is a free function of $K$. Some authors have
chosen $F(K)$ in the following forms: (i) $F(K)=\gamma(-K)^{n}$
\cite{Zlos,Junt}, (ii)
$F(K)=\gamma\sqrt{-K}+\sqrt{\frac{3K}{\beta}}~\ln(-K)$
\cite{Meng1,Meng2}. Here we may choose another form of $F(K)$ for
our next calculations in simplified form as
$F(K)=\frac{2}{\beta}~K(1-\epsilon~ K)$, where $\epsilon$ is a
constant. So solving equation (13), we obtain the expression of
$H^{2}$ in terms of redshift $z$ in the following:
\begin{equation}
H^{2}(z)=\frac{M}{3\sqrt{3\epsilon\beta}}\left[-3k(1+z)^{-2}+8\pi
G \rho_{m0}(1+z)^{3(1+w_{m})}+8\pi G
\rho_{ch0}\left\{A_{s}+(1-A_{s})(1+z)^{3(1+A)(1+\alpha)}
\right\}^{\frac{1}{1+\alpha}}   \right]^{\frac{1}{2}}
\end{equation}

Now defining the dimensionless parameters $\Omega_{m0}=\frac{8\pi
G\rho_{m0}}{3H_{0}^{2}}$~, $\Omega_{ch0}=\frac{8\pi
G\rho_{ch0}}{3H_{0}^{2}}$~, $\Omega_{k0}=\frac{k}{H_{0}^{2}}$ and
$\Omega_{EA}=\frac{M}{3H_{0}\sqrt{\epsilon\beta}}$, we obtain the
form of $H(z)$:
\begin{equation}
H(z)=H_{0}\sqrt{\Omega_{EA}}\left[-\Omega_{k0}(1+z)^{-2}+
\Omega_{m0}(1+z)^{3(1+w_{m})}+
\Omega_{ch0}\left\{A_{s}+(1-A_{s})(1+z)^{3(1+A)(1+\alpha)}
\right\}^{\frac{1}{1+\alpha}}   \right]^{\frac{1}{4}}
\end{equation}
Due to the above solution, the equation (13) gives the following
relation:
\begin{equation}
\sqrt{\Omega_{EA}}~\left[\Omega_{m0}+\Omega_{ch0}-\Omega_{k0}\right]=1.
\end{equation}

\section{\normalsize\bf{Observational Data Analysis Tools}}

In this section, we shall investigate some bounds of the
parameters of the modified Chaplygin gas (MCG) in Einstein-Aether
gravity by observational data fitting. The parameters are
determined by observed Hubble data (OHD), BAO, CMB and SNe data
analysis
\cite{Meng1,Meng2,Wu1,Paul,Paul1,Paul2,Paul3,Deb1,Deb2,Deb3,Deb4}.
We shall use the $\chi^{2}$ minimization technique (statistical
data analysis) from Hubble-redshift data set to get the
constraints of
the parameters of MCG model in Einstein-Aether gravity.\\

\[
\begin{tabular}{|c|c|c|}
\hline
  ~~~~~~$z$ ~~~~& ~~~~$H(z)$ ~~~~~& ~~~~$\sigma(z)$~~~~\\
  \hline
  0 & 73 & $\pm$ 8 \\
  0.1 & 69 & $\pm$ 12 \\
  0.17 & 83 & $\pm$ 8 \\
  0.27 & 77 & $\pm$ 14 \\
  0.4 & 95 & $\pm$ 17.4\\
  0.48& 90 & $\pm$ 60 \\
  0.88 & 97 & $\pm$ 40.4 \\
  0.9 & 117 & $\pm$ 23 \\
  1.3 & 168 & $\pm$ 17.4\\
  1.43 & 177 & $\pm$ 18.2 \\
  1.53 & 140 & $\pm$ 14\\
  1.75 & 202 & $\pm$ 40.4 \\ \hline
\end{tabular}
\]
{\bf Table 1:} The observed Hubble parameter $H(z)$ and the
standard error $\sigma(z)$ for different values of redshift $z$.

\subsection{\normalsize\bf{Analysis with Observed Hubble Data (OHD)}}

We analyze the MCG model in Einstein-Aether gravity using observed
value of Hubble parameter data (OHD)\cite{Stern,Zhang1} at
different redshifts consists of twelve data points. The observed
values of Hubble parameter $H(z)$ and the standard error
$\sigma(z)$ for different values of redshift $z$ are listed in
Table 1. The $\chi^{2}$ statistics for OHD is give as follows:
\begin{equation}
{\chi}_{OHD}^{2}=\sum\frac{(H(z)-H_{obs}(z))^{2}}{\sigma^{2}(z)}
\end{equation}

where $H(z)$ and $H_{obs}(z)$ are respectively the theoretical and
observational values of Hubble parameter at different redshifts
and $\sigma(z)$ is the corresponding error which is given in table
1. We consider the present value of Hubble parameter $H_{0}$ = 72
$\pm$ 8 Kms$^{-1}$ Mpc$^{-1}$. Here we shall determine two
parameters of MCG model out of 3 parameters $A,~B,~\alpha$ by
fixing any one parameter from minimizing the above distribution
${\chi}_{OHD}^{2}$. There are other parameters of the model say
$\Omega_{m0},~\Omega_{k0},~\Omega_{ch0},\Omega_{EA},~w_{m}$.
Fixing the one parameter $\alpha$ of MCG model, the relation
between the other parameters $(A,B)$ can be determined by the
observational data. Now for OHD analysis, $\chi^{2}_{OHD}$ is
minimized for best fit values of $A=0.238303$ and $B=0.18176$ and
the minimum value of $\chi^{2}_{OHD}=7.08613$ where we have
assumed $\alpha=0.1$. We also plot the graph for different
confidence levels (66\%, 90\%, 99\%) in figure 1.

\begin{figure}

\includegraphics[height=3.0in]{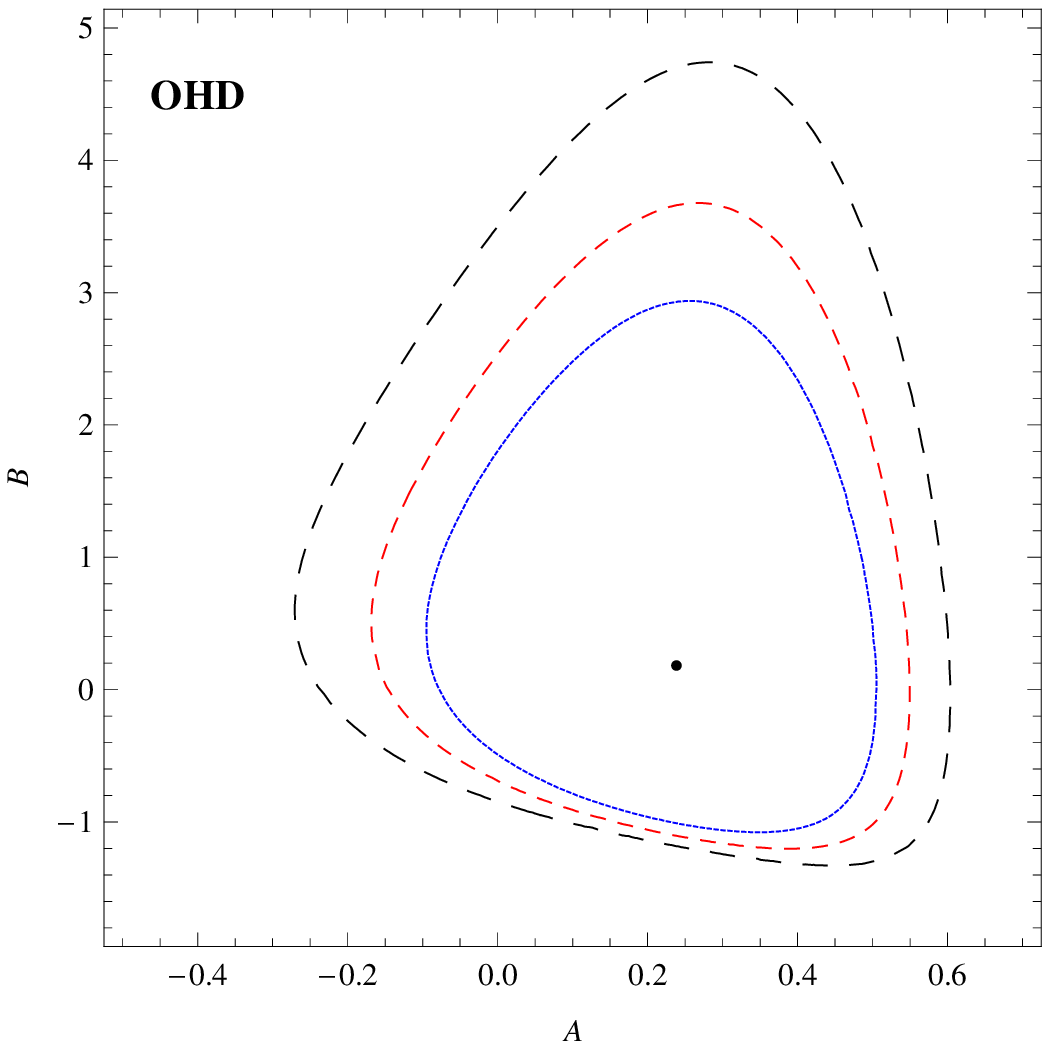}~~~~
\includegraphics[height=3.0in]{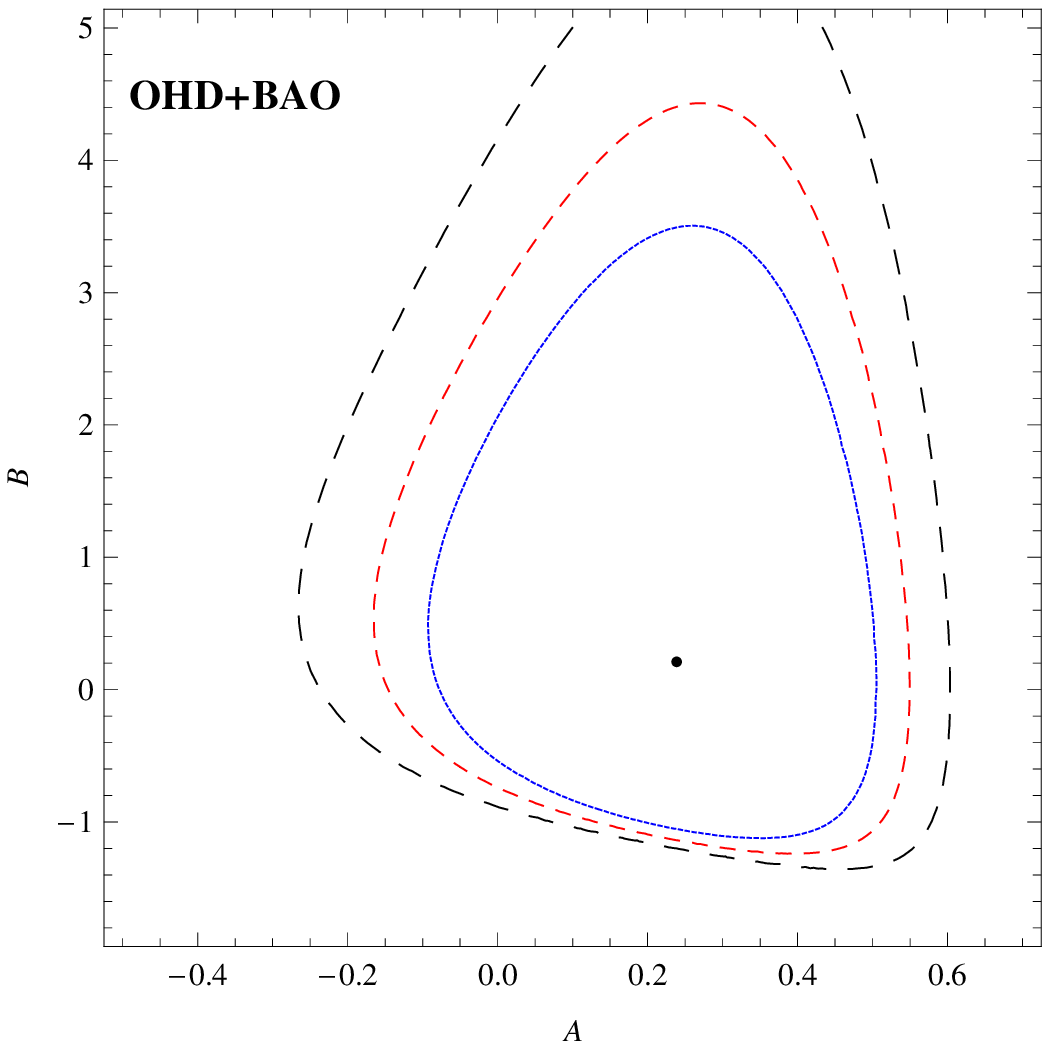}
\vspace{2mm}
~~~~~~Fig.1~~~~~~~~~~~~~~~~~~~~~~~~~~~~~~~~~~~~~~~~~~~~~~~~~~~~~~~~~~~~~~~~Fig.2~~\\
\vspace{4mm}

\vspace{.2in} Figs. 1 and 2 show the variations of $A$ with $B$
for $\alpha=0.1$ for OHD and OHD+BAO analysis respectively for
different confidence levels say 66\% (solid, blue), 90\% (dashed,
red) and 99\% (dashed, black) contours. \vspace{0.2in}
\end{figure}

\subsection{\normalsize\bf{Analysis with OHD+BAO}}

Another constraint is from the Baryonic Acoustic Oscillations
(BAO) traced by the Sloan Digital Sky Survey (SDSS). The BAO peak
parameter value has been proposed by Eisenstein et al
\cite{Eisenstein}. Here we examine the parameters $A$ and $B$ for
MCG gas model from the measurements of the BAO peak for low
redshift (with range $0<z<0.35$) using standard $\chi^{2}$
analysis. The BAO peak parameter may be defined by \cite{Meng1}
\begin{equation}
{\cal A}=\frac{\sqrt{\Omega_{m}}}{\left\{\Omega_{k}
E(z_{1})\right\}^{1/3}}
~\left[\frac{1}{z_{1}}~\sinh\left\{\sqrt{\Omega_{k}}
\int_{0}^{z_{1}} \frac{dz}{E(z)}\right\}\right]^{\frac{2}{3}}
\end{equation}
where $E(z)=H(z)/H_{0}$ may be called the normalized Hubble
parameter, the redshift $z_{1}=0.35$ is the typical redshift of
the SDSS. The value of the parameter ${\cal A}$ for the universe
is given by ${\cal A}=0.469\pm 0.017$ using SDSS data
\cite{Eisenstein}. Now the $\chi^{2}$ function for the BAO
measurement can be written as
\begin{equation}
\chi^{2}_{BAO}=\frac{({\cal A}-0.469)^{2}}{(0.017)^{2}}
\end{equation}

Now the total joint data analysis of BAO with OHD for the
$\chi^{2}$ function may be defined by
\begin{equation}
\chi^{2}_{total}=\chi^{2}_{OHD}+\chi^{2}_{BAO}
\end{equation}

According to OHD+BAO joint analysis the best fit values of $A$ and
$B$ are $A=0.238695$ and $B=0.209932$ with $\chi^{2}$ minimum is
7.07842. Finally we draw the contours $B$ vs $A$ for the 66\%,
90\% and 99\% confidence limits depicted in figure 2.

\subsection{\normalsize\bf{Analysis with OHD+BAO+CMB}}

In addition to OHD and BAO analysis, we use the Cosmic Microwave
Background (CMB) shift parameter. The CMB shift parameter (CMB
power spectrum first peak) is defined by
\cite{Bond,Efstathiou,Nessaeris}

\begin{equation}
{\cal
R}=\frac{\sqrt{\Omega_{m}}}{\sqrt{\Omega_{k}}}~\sinh\left[\sqrt{\Omega_{k}}
\int_{0}^{z_{2}} \frac{dz}{E(z)}\right]
\end{equation}

where $z_{2}$ is the value of redshift at the last scattering
surface. From 7 year WMAP data \cite{Komatsu}, the value of the
parameter has obtained as ${\cal R}=1.726\pm 0.018$ at the
redshift $z_{2}=1091.3$. Now the $\chi^{2}$ function for the CMB
measurement can be written as
\begin{equation}
\chi^{2}_{CMB}=\frac{({\cal R}-1.726)^{2}}{(0.018)^{2}}
\end{equation}

Now when we consider OHD, BAO and CMB analysis together, the total
joint data analysis (OHD+BAO+CMB) for the $\chi^{2}$ function may
be defined by
\begin{equation}
\chi^{2}_{TOTAL}=\chi^{2}_{OHD}+\chi^{2}_{BAO}+\chi^{2}_{CMB}
\end{equation}

Now the best fit values of $A$ and $B$ with $\chi^{2}$ for joint
analysis of BAO and CMB with OHD observational data support the
theoretical range of the parameters. The best fit values are
$A=0.239018$ and $B=0.240047$ with the minimum value of
$\chi^{2}=7.07086$. The 66\%, 90\% and 99\% contours for $A$ and
$B$ are plotted in figure 3.

\begin{figure}

\includegraphics[height=3.0in]{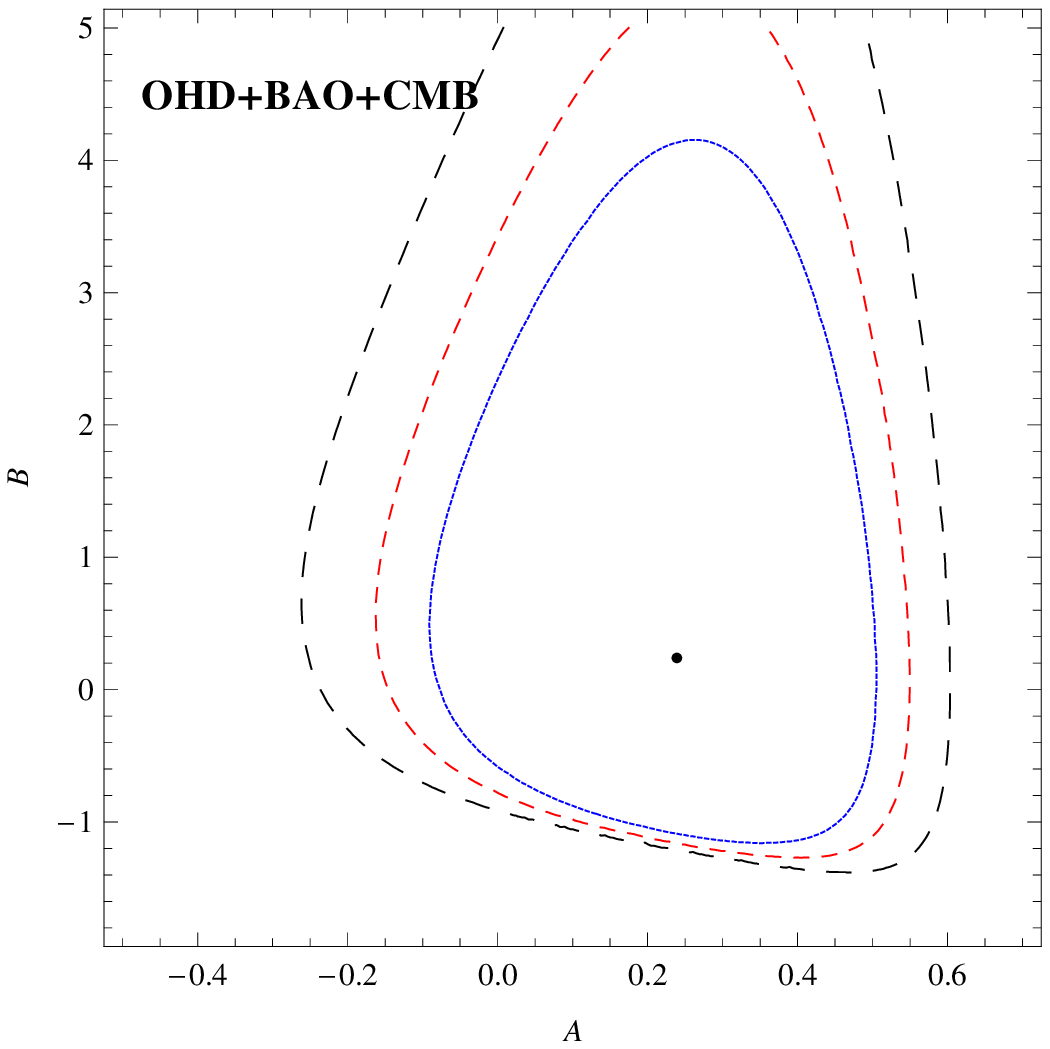}~~~~
\includegraphics[height=3.0in]{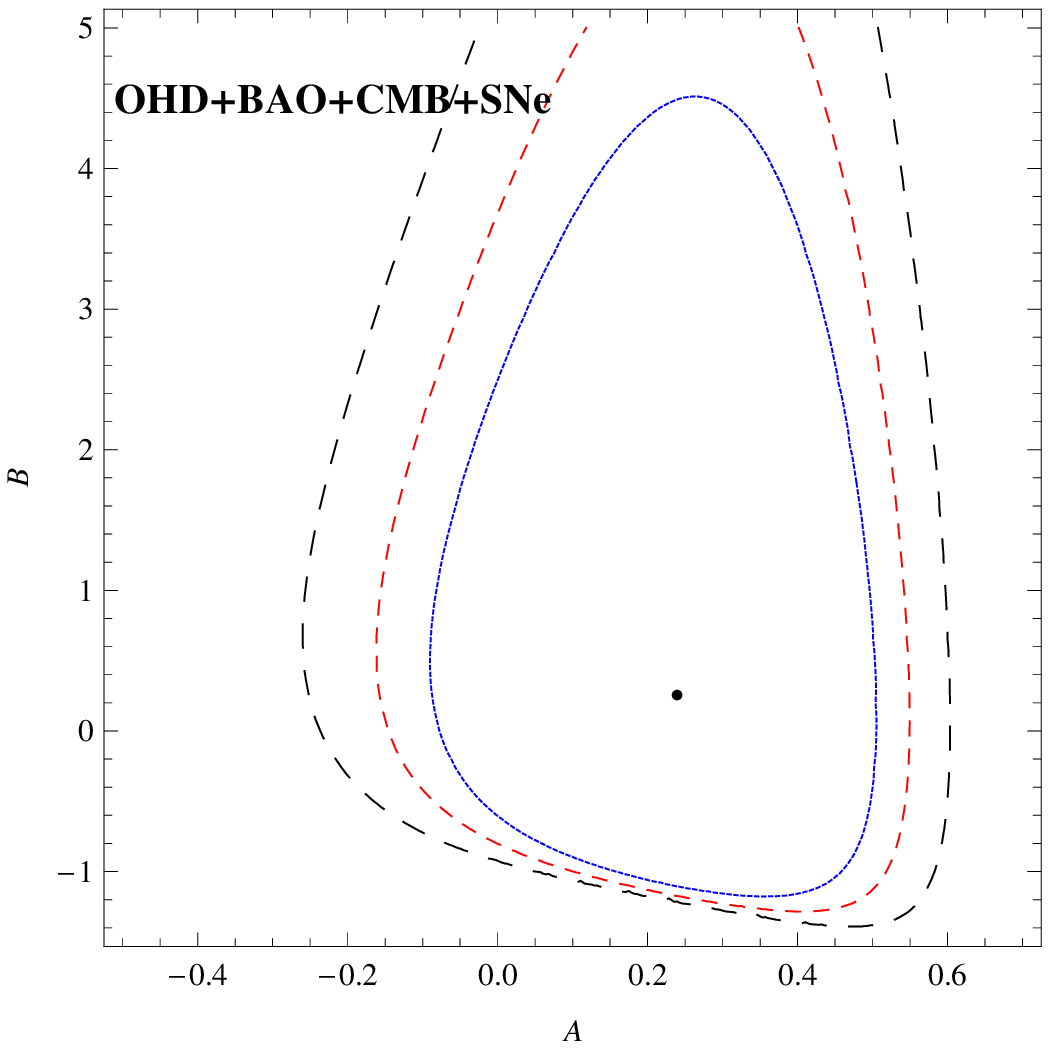}
\vspace{2mm}
~~~~~~Fig.3~~~~~~~~~~~~~~~~~~~~~~~~~~~~~~~~~~~~~~~~~~~~~~~~~~~~~~~~~~~~~~~~Fig.4~~\\
\vspace{4mm}

\vspace{.2in} Figs. 3 and 4 show the variations of $A$ with $B$
for $\alpha=0.1$ for OHD+BAO+CMB and OHD+BAO+CMB+SNe analysis
respectively for different confidence levels say 66\% (solid,
blue), 90\% (dashed, red) and 99\% (dashed, black) contours.
\vspace{0.2in}
\end{figure}

\subsection{Redshift-Magnitude Observations from Supernovae Type Ia:\\
Analysis with OHD+BAO+CMB+SNe Ia}

The observations of type Ia supernovae (SNe Ia) provide an
excellent tool for probing the expansion history of the universe.
The main evidence for the existence of dark energy is provided by
the Supernova type Ia experiments \cite{Perlmutter,Riess,Riess1}.
The type Ia observations directly measure the distance modulus of
a Supernovae and its redshift $z$ \cite{Riess2,Kowalaski}. Now,
take recent observational data (including SNe Ia) consists of 557
data points and belongs to the Union2 sample \cite{Amanullah}.
From the type Ia observations, the luminosity distance determines
the dark energy density. The luminosity distance $d_{L}(z)$ is
defined by
\begin{equation}
d_{L}(z)=\frac{(1+z)}{\sqrt{\Omega_{k}}}~\sinh\left[\sqrt{\Omega_{k}}~\int_{0}^{z}\frac{dz'}{E(z')}\right]
\end{equation}

and the distance modulus $\mu(z)$ for Supernovas is given by
\begin{equation}
\mu(z)=5\log_{10}
\left[\frac{d_{L}(z)/H_{0}}{1~\text{MPc}}\right]+25
\end{equation}

The $\chi^{2}$ function for SNe Ia is given by
\begin{equation}
{\chi}_{SNe}^{2}=\sum\frac{(\mu(z)-\mu_{obs}(z))^{2}}{\sigma^{2}(z)}
\end{equation}

where $\mu_{obs}(z)$ is observational value of distance modulus
parameter at different redshifts and $\sigma(z)$ is the
corresponding error. In this work, we take Union2 data set
consisting of 557 supernovae data. Now we consider four
cosmological tests together, the total joint data analysis
(Stern+BAO+CMB+SNe) for the $\chi^{2}$ function may be defined by
\begin{equation}
\chi^{2}_{TOTAL}=\chi^{2}_{OHD}+\chi^{2}_{BAO}+\chi^{2}_{CMB}+\chi^{2}_{SNe}
\end{equation}

From the joint analysis, we found the minimum value of $\chi^{2}$
and which is 7.06716. The best fit values of the parameters are
$A=0.239158$ and $B=0.255814$. The confidence contours are drawn
in figure 4. The best fit value of distance modulus $\mu(z)$ for
our theoretical model and the Supernova type Ia union2 sample are
drawn in figure 5 for our best fit values of $A$ and $B$. From the
curves, we see that the theoretical MCG model in Einstein-Aether
gravity is in agreement with the union2 sample data.

\begin{figure}

\includegraphics[height=3.0in]{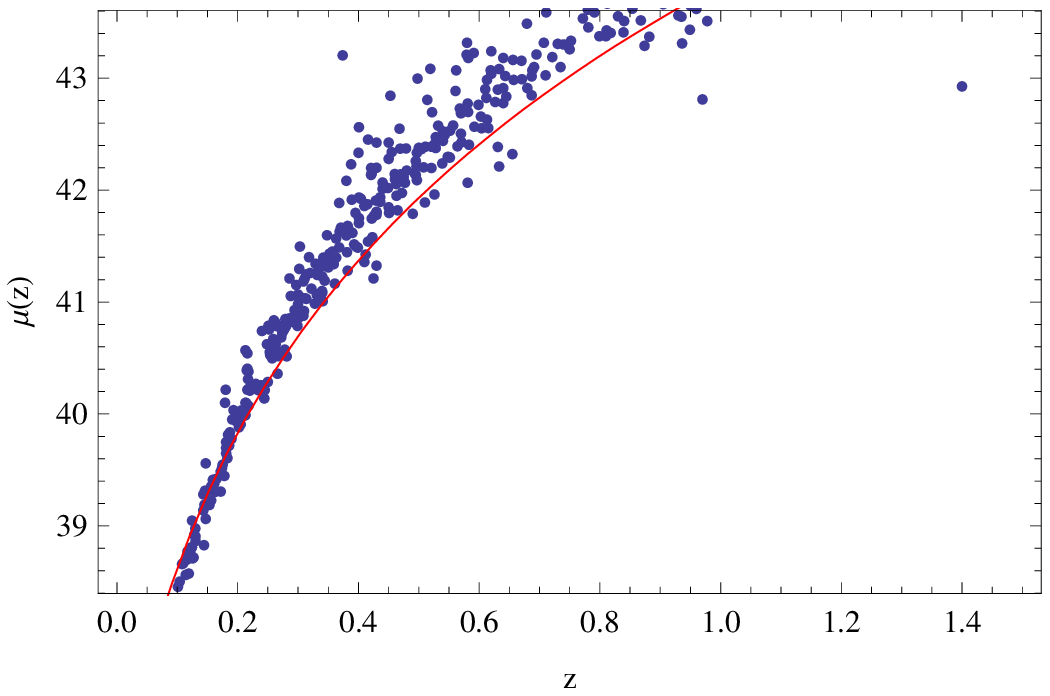}~~~~\\
\vspace{2mm}
~~~~~~~Fig.5~\\
\vspace{4mm}

\vspace{.2in} Fig.5 shows the variation of distance modulus
$\mu(z)$ vs redshift $z$ for our model (solid line) and the Union2
sample (dotted points). \vspace{0.2in}
\end{figure}

\section{\normalsize\bf{Discussions and Concluding Remarks}}

We have assumed FRW model of the universe in Einstein-Aether
gravity filled with dark matter and Modified Chaplygin gas (MCG)
type dark energy. Dark matter has the equation of state parameter
$w_{m}$, which is small. We assumed the dark matter and dark
energy separately conserved and hence we found the solutions in
this gravity. Since $F(K)$ is a free function of $K$, so we have
chosen quadratic form of $F(K)$ for simplicity of the calculation.
Defining dimensionless parameters, we present the Hubble parameter
in terms of some unknown parameters and observational parameters
with the redshift $z$. From observed Hubble data (OHD) set (12
points), we have obtained the bounds of the arbitrary parameters
($A,B$) of MCG by minimizing the $\chi^{2}$ test where we have
chosen $\alpha=0.1$. The minimum values of the parameters are
$A=0.238303$ and $B=0.18176$ for OHD analysis. Next due to joint
analysis of BAO and CMB observations, we have also obtained the
best fit values and the bounds of the parameters ($A,B$). The best
fit values of the parameters (i) for OHD+BAO are $A=0.238695$ and
$B=0.209932$ and (ii) for OHD+BAO+CMB are $A=0.239018$ and
$B=0.240047$. We have also taken type Ia supernovae data set
(union 2 data set with 557 data points). Next due to joint
analysis with SNe, we have obtained the best fit values of the
parameters $(A,B)$. The best fit values of the parameters for
OHD+BAO+CMB+SNe are $A=0.239158$ and $B=0.255814$. The best-fit
values and bounds of the parameters are obtained by 66\%, 90\% and
99\% confidence levels for OHD, OHD+BAO, OHD+BAO+CMB and
OHD+BAO+CMB+SNe joint analysis in figures 1-4. The distance
modulus $\mu(z)$ against redshift $z$ for our theoretical MCG
model in Einstein-Aether gravity have been tested for the best fit
values of the parameters and the observed SNe Ia union2 data
sample and drawn in figure 5. The observations do in fact severely
constrain the nature of allowed composition of matter-energy by
constraining the range of the values of the parameters for a
physically viable MCG in Einstein-Aether gravity model. \\

\section*{Acknowledgements}

The author is thankful to Institute of Theoretical Physics,
Chinese Academy of Science, Beijing, China for providing TWAS
Associateship Programme under which part of the work was carried
out. Also UD is thankful to CSIR, Govt. of India for providing
research project grant (No. 03(1206)/12/EMR-II). \\

\end{document}